# The Oxford Insights Government AI Readiness Index (GARI): An Analysis of its Data and Overcoming Obstacles, with a Case Study of Iraq

Ahmed Shaker Alalaq

Iraq\University of Kufa

ahmed.alallaq@uokufa.edu.iq

https://orcid.org/0000-0002-8033-9172

**summary**

The research examines the "Government AI Readiness Index" (GARI) issued by Oxford, analyzing data related to governmental preparedness for adopting artificial intelligence across different countries. It highlights the evaluation criteria for readiness, such as technological infrastructure, human resources, supportive policies, and the level of innovation.

The study specifically focuses on Iraq, exploring the challenges the Iraqi government faces in adopting and implementing AI technologies. It addresses economic, social, and political barriers hindering this transition and provides concrete recommendations to overcome these obstacles. Through Iraq's case study, the research aims to offer insights into improving public-private sector collaboration to enhance the effective use of AI in governance and public administration.

The study emphasizes the importance of investing in education, training, and capacity building to develop the necessary workforce, enabling countries to reap the benefits of AI and improve the efficiency of government services.

Keywords (Artificial Intelligence, Iraq, Index, Data, Results, Laws)

**Introduction:**

The rise of artificial intelligence (AI) and the subsequent revolution in digital technologies have presented us with a vast new landscape. This requires a broad perspective, an open mind, and a willingness to embrace the new ideas and technological advancements that are emerging on the horizon. We are facing a significant and serious responsibility if we want our universities to occupy a prominent position and our research centers to achieve advanced rankings on the global stage. This is the ambition we strive to achieve in the near future.

First and foremost, it must be acknowledged that there is a significant gap and hesitation in the adoption of AI principles and concepts. We need to take serious and concrete steps in this field. The first step that governments must take is to update and develop their teaching staff and make them receptive to this technology. They must also deal with it seriously as an integral part of achieving quality and scientific development, of which the most prominent features are sound scientific research and publication at the level of reputable international repositories.

In this research, we attempt to provide an initial reading of the Oxford Insights index regarding the readiness of governments around the world to embrace the AI revolution. We will also identify the most prominent obstacles that hinder the achievement of this technology, which has become a central and clearly defined focus within all educational institutions. We will also try to identify the strengths of these institutions and the extent of our interaction as an educational community with the products of modern technology and digitization, including program interfaces, applications, and accredited scientific documentation elements.

**Research Problem:** The study sheds light on an initial reading of the Oxford Insights index from an Arab perspective and examines the data it contains according to the regions classified within the index's terms. The study also aims to identify the most prominent obstacles to the development and modernization of the AI work system and to show the extent of the differences between countries around the world based on political, economic, and technological factors. We also aim to identify and study the case of Iraq, which was included in the aforementioned index, and to clarify, albeit in a simplified manner, the most prominent solutions that help, in one way or another, to overcome the decline of the digital learning system and the reliance on AI tools.

**Study Objectives:** The study aims to achieve the following objectives:

1- To identify the main contents of the British Oxford Insights index.
2- To conduct a comprehensive survey of the rankings of the countries included in the index using charts according to the ranking of each country.
3- To identify the cases of decline and fear that have affected most countries of the world regarding the use of artificial intelligence technologies.
4- To present the case of Iraq and to show the extent to which its institutions are able to deal with and respond to the data and outputs of artificial intelligence.
5- To clarify a set of outputs and proposals that would enhance Iraq's position globally and regionally regarding the adoption of artificial intelligence.

1. **A Reading of the Oxford Insights AI Readiness Index**

The Government AI Readiness Index (AI Readiness Index) is a global index that measures the extent to which governments are prepared to use artificial intelligence (AI) effectively to improve public services and outcomes for citizens. The index was developed by Oxford Insights, a British global advisory firm, in collaboration with the International Development Research Centre (IDRC), a Canadian non-profit organization.

Not long ago, Oxford Insights published a report entitled The Government AI Readiness Index 2023, which ranked 193 countries in terms of their readiness and acceptance of AI technologies. The report summary, which was published on the Oxford Insights website, stated that 2023 witnessed a significant and clear interest from everyone, as evidenced by the holding of conferences, seminars, and workshops that clarified the nature of AI tools. Many countries have explicitly acknowledged that AI has played a clear role in overcoming obstacles to scientific progress and that they are striving to integrate many of the foundations of this intelligent revolution into the public services sector in their countries. The question that remains is: To what extent are governments prepared to apply AI in the delivery of public services to their citizens?

The index focuses on three axes:

1- Government Environment: This axis measures the existence of a government AI strategy, the adoption of supporting regulations and legislation, and the ability to use AI technologies to overcome public service challenges.
2- Technology: This axis measures the growth of the technology ecosystem and its focus on AI, including the presence of AI startups, investment in research and development, and up skilling the workforce.
3- Data and Infrastructure: This axis measures the availability of data, digital skills, and the ICT infrastructure needed to develop and use AI. It essentially assesses the capacity of these infrastructures and human resources and the speed at which they can adopt the outputs of AI software.

The report concluded that the government axis has seen a decline in its productivity and interaction with AI tools on an annual basis, due to the reluctance of some of these areas to adopt AI applications. However, this year has seen a noticeable shift, as half of the AI strategies launched or announced come from low- and middle-income countries, which is the opposite of what was expected.

As for the technology axis, there is a disparity between high-income countries and other countries, although some large middle-income economies outperform their weight. High-income countries score significantly higher than countries in any other income group on the technology sector pillar, with the gap between high-income and upper-middle-income countries being larger than the gaps between all other income groups combined in some cases. However, large middle-income countries like Malaysia outperform their income groups and rank among the top 50 countries globally on this particular axis.

The infrastructure axis was also discussed, and while generative AI "bodes well" for low-income countries, the lack of a solid foundation in data and infrastructure could lead to reliance on foreign technology, which could lead to obstacles such as language barriers. For example . (**oxfordinsights.com**).

The index focused on a set of real dimensions that are in complete alignment with AI tools, namely:

- Data and privacy
- Cyber security
- Ethics framework for dealing with smart tools
- Legal legislation to adapt to digital business models
- Encouraging government investment in emerging technologies
- Use of ICT and government efficiency
- Online services
- Trust in government websites and applications
- And other indicators. (**Aldane, 2022**) .

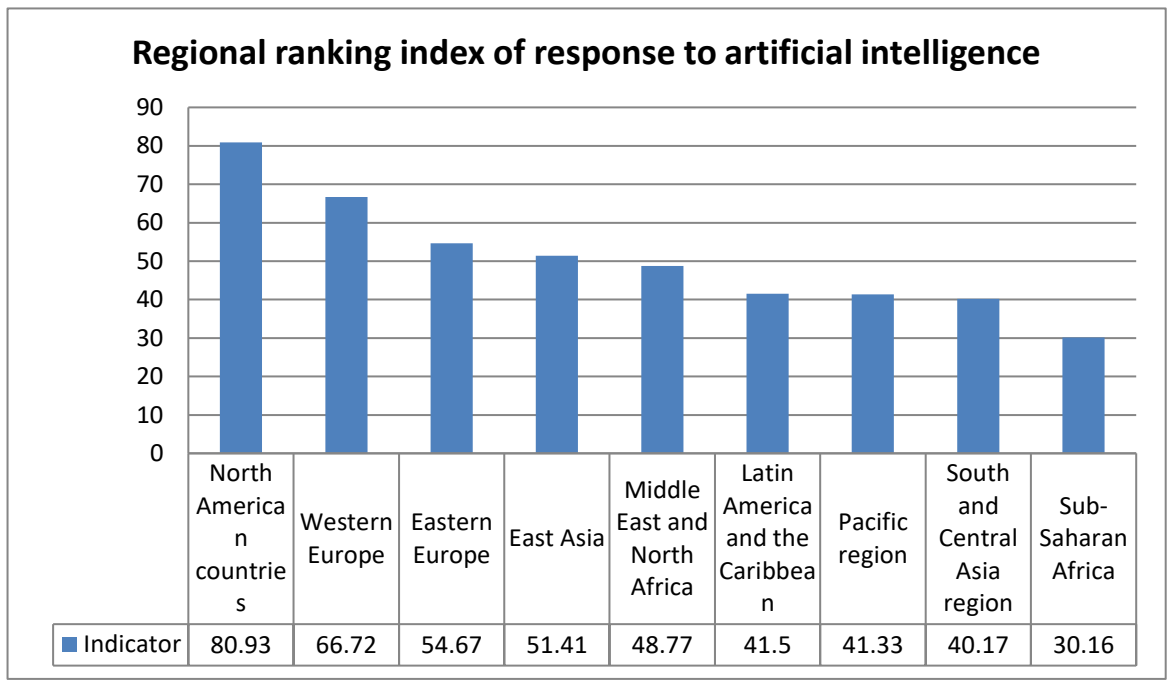

**Regional ranking index of response to artificial intelligence**

| | North American countries | Western Europe | Eastern Europe | East Asia | Middle East and North Africa | Latin America and the Caribbean | Pacific region | South and Central Asia region | Sub-Saharan Africa |
|---|---|---|---|---|---|---|---|---|---|
| Indicator | 80.93 | 66.72 | 54.67 | 51.41 | 48.77 | 41.5 | 41.33 | 40.17 | 30.16 |

At first glance, looking at these indicators and results, and ranking countries according to their scores gives a clear impression that the political, economic, and military capabilities, the growth of infrastructure strength, and the increasing intensity of work to update the digital revolution system in North American countries, led by the United States of America, come from the desire of those countries to be the first to embrace and adopt the technology revolution. This is a matter that has been imposed by the circumstances we mentioned at the beginning of our talk.

The dominance of North American countries, led by the United States of America with an index of (84.80), followed by Canada with an index of (77.07), at the top of the index is a natural situation given that these countries possess all the qualifications that make them at the top of the pyramid. From a political point of view, North American countries generally enjoy tremendous political capabilities that make them able to keep up with and contain all the events and developments surrounding all countries of the world. As it is known, the stability of the political situation in any country makes that country able to reach a state of development and progress in all fields and makes its individuals capable of being a role model if those governments provide what they should provide for their people.

Most of the artificial intelligence applications were the result of American companies, and there are laws that require all those companies to work diligently to update the system of those programs. In addition, American governments have been and continue to be sponsors and supporters of such technological activities by providing financial resources for them and enacting laws that encourage technical, engineering, and programming work. In many cases, those governments purchase the products of the work of those research centers that produce intelligent programs.

Moreover, the economic reality of those countries is known to be strong and powerful to the extent that they are among the countries that provide financial and logistical assistance to many countries of the world during crises. This necessarily has a positive impact on the reality of education and learning, and thus provides all the needs for the progress and development of their educational institutions with huge financial resources and expenditures that often exceed the expenditures provided for other aspects of life such as health, the army, or other service sectors. (**Nzobonimpa, 2023**).

These indicators naturally have a positive or negative impact on the reality of receiving the results of the education revolution and smart technologies. We should not forget that most of those software and artificial intelligence tools are made by those countries, so it is natural for those countries to be at the top of the artificial intelligence revolution response index, with what they have of highly advanced educational institutions, mechanisms, devices, and technology programs.

As for the Middle East artificial intelligence acceptance index, it came in fifth place globally with a score of (48.77). This is a situation that imposes itself based on the political and economic data that most countries in the Middle East and North Africa enjoy. The ranking of countries in this region is as follows:

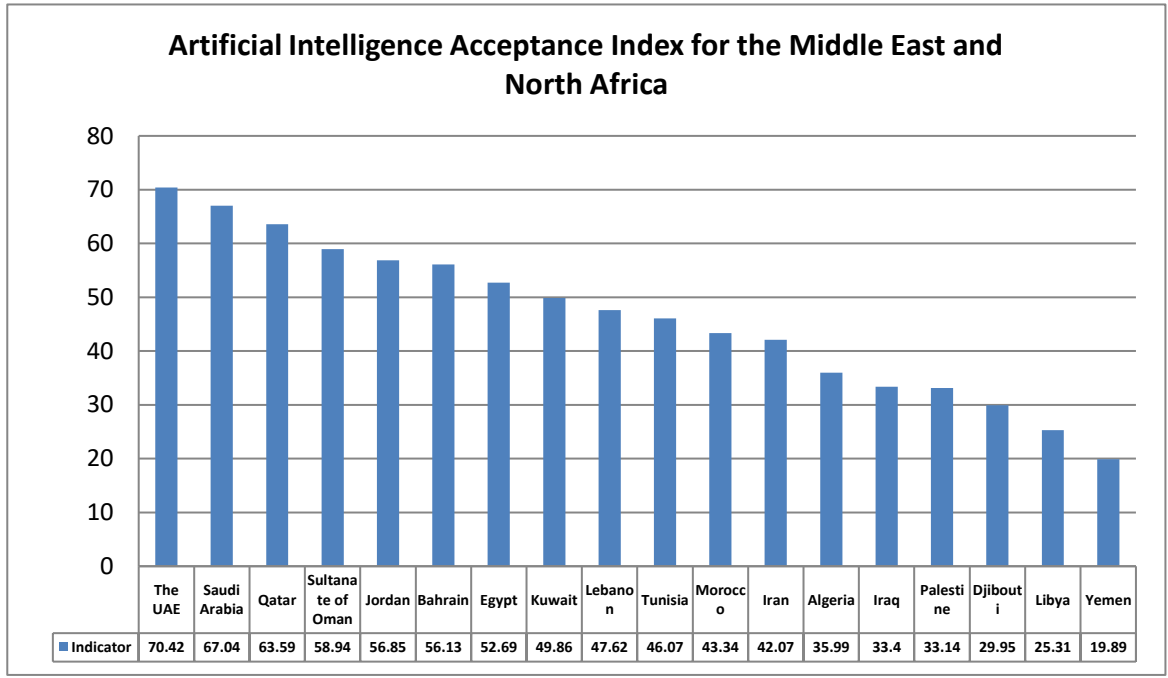

Artificial Intelligence Acceptance Index for the Middle East and North Africa

| | The UAE | Saudi Arabia | Qatar | Sultanate of Oman | Jordan | Bahrain | Egypt | Kuwait | Lebanon | Tunisia | Morocco | Iran | Algeria | Iraq | Palestine | Djibouti | Libya | Yemen |
|---|---|---|---|---|---|---|---|---|---|---|---|---|---|---|---|---|---|---|
| Indicator | 70.42 | 67.04 | 63.59 | 58.94 | 56.85 | 56.13 | 52.69 | 49.86 | 47.62 | 46.07 | 43.34 | 42.07 | 35.99 | 33.4 | 33.14 | 29.95 | 25.31 | 19.89 |

Iraq is ranked 133rd globally and 14th in the Middle East and North Africa region with an index of (33.40). A close, analytical look at Iraq's ranking among countries in the world and the region in which it is located shows that Iraq has occupied a good position compared to other countries in the world that have preceded Iraq with many capabilities at all levels.

What is required now is to intensify efforts and accelerate the pace for Iraq to reach higher ranks in the ability to use artificial intelligence applications. There are plans and studies that have been put in place to achieve this goal.

The fact that Middle Eastern and North African countries, including Arab countries or countries that are classified as having little experience in the smart technology revolution, have low ranks, except for some countries such as the Emirates and Saudi Arabia, in the index of receiving artificial intelligence tools worldwide is due to the lack of interest in this technology and the lack of serious consideration of the importance of these programs and their reliance on them to overcome obstacles to development and work quality issues, not only in educational institutions, but in all sectors of life.

In addition, all strategic plans at the government level are immature in idea and unclear in features, with a severe shortage of data on the overall phenomenon of dealing with this revolution.

So far, the belief prevails that such technologies are not useful in front of the human mind. Yes, we do not deny the role of the human mind that produced such tools in the continuity and updating of this technological technology, but it is also necessary to pay increasing attention to developing the capabilities of that human mind to absorb the outputs of artificial intelligence technology.

I believe that there is a clear fear among many that many specialists and workers in certain fields will be replaced by artificial intelligence technology, and there are practical indicators using robots that work with artificial intelligence. In fact, the matter has reached the point where many companies and institutions have abandoned most of their employees.

We must also point out that there is a fear on the other hand among many governments of these countries or opinion leaders and decision-makers from embarking on this experiment, which they see as very difficult and its results may be, according to some of them, the opposite, casting a shadow on the work of this institution or that.

This fear is due to the lack of experienced and skilled people in some of those institutions who are able to deal with artificial intelligence technology. They also see that this technology is still new and shrouded in mystery, so they are not willing to risk the reputation of their institutions, especially the academic ones, and to throw these programs into the behavior of learning patterns and make them a vital and main part of their institutions. Therefore, all of these outputs may be subject to espionage by the centers, individuals, or companies that produce these tools.

In conclusion, it can be said that using such a precise global index allows governments and companies to have a better, broader, and more comprehensive understanding of their ability to absorb artificial intelligence tools and motivate their human bases to improve the work of their institutions in preparation for accepting the renewed artificial intelligence technology.

## 2. What do we need to keep up with the artificial intelligence revolution?

First, we must acknowledge that there is a significant and clear lack of knowledge among employed elites working in many educational institutions around the world regarding artificial intelligence and API applications. This is due to a lack of experience, weak or even scarce technological equipment, and a lack of familiarity with them for years. Many professors do not know how to use their personal computers or how to guide their students to benefit from the technical capabilities that some scientific institutions, such as universities and research centers, may provide.

Let us be honest and admit that if it were not for some of the emergency crises that emerged, such as the Corona crisis, and the forced use of computers and distance learning technologies by everyone, some of us would not have been able to overcome the "technology complex", if I may say so. This complex has accompanied many professors and students for many years before the Corona pandemic crisis.

Now we are facing a huge task, which is how to overcome this complex and how to receive the outputs of artificial intelligence technology.

Governments and decision-makers are required to develop a set of firm decisions and recommendations that will allow us to overcome the weaknesses we suffer from regarding the use of computing technologies.

Now, senior leaders are directly responsible for making decisions and implementing them immediately and decisively if we want to move forward in updating educational systems in line with what international universities have reached, even if it is to a small extent, depending on the economic capabilities of each country.

The economic aspect is essential in this regard. It is necessary to provide and allocate huge sums of money to improve the work of digital devices and information communication tools, including internet networks, paid programs and applications. Many of us now rely on free programs and applications. However, as is known, these applications do not provide the wide and unlimited capabilities and services that paid applications do. I mean those programs that are concerned with generating texts, creating scientific analytical outputs, charts and presentations, accurate translation programs, anti-spyware programs, and other cyber security programs, as well as other scientific programs.

Increasing financial allocations will provide tremendous opportunities for all educational elites, from professors to undergraduate and graduate students, to use these APIs for free, paid for by the government. Thus, we will be able to keep up with countries that have made great strides in this field.

This point, i.e. financial allocations, is also beneficial for lecturers and is an incentive for them to move forward in learning artificial intelligence programs if the government provides financial incentives for anyone who produces a program or application or works to update any of these APIs that may benefit and enhance the process of updating the educational process in the institutions of this or that country.

It can also be said that one of the most prominent things that some scientific institutions lack is the lack of technical devices, or they are almost non-existent in many of them. There are many professors who do not own a personal computer or their institution cannot provide them with one. I think it is best for these institutions to adopt a work plan to provide these devices at a subsidized cost, and to impose on the professor the use of the computer and e-learning technologies during the lecture.

Let us assume that there are about 90% of professors in any college, institute, or research center who use the computer professionally. We will see a quantum leap in the field of using communication technologies between the professor and the student. In such cases, the professor will be forced to discover for himself what digital technologies have produced in terms of advanced applications and programs. Thus, he will transfer this experience to his students who will deal with this new technology with some seriousness. Then we will see a noticeable and serious increase in the indicators of digital education, which includes artificial intelligence technologies.

Here, the top government leaders are directly responsible for following up on the lecturer regarding the extent of his use of this technology, his long-term readiness, and the extent to which his students accept the reality of this technology. The matter should be imposed in an encouraging manner, with the provision of a set of material and moral incentives. Thus, we will see amazing results.

Another thing I see fit in this regard is to intensify the holding of conferences, workshops, and lectures that educate on the use of e-learning technologies and artificial intelligence programs. Such training courses help, in one way or another, to promote the culture of using computers and useful programs. There are indicators that have appeared in regional countries, and by following such paths, they have emerged from nothing to become leaders in the use of digital technologies and educational software.

The matter is very easy, as all our institutions have a good number of professors who specialize in the use of technical programs. Through them, we can train or teach a wide range of professors with little experience in this field, and we can reap the results of this process within an academic year or two.

Another issue I would like to point out is that those in charge should intensify their efforts to implement the series of programs and plans in the best possible way. They should not only focus on quantity, but also on the quality of the outputs. We are all morally and legally responsible for achieving these lofty goals in order to improve the work of our educational institutions and research centers and to keep up with the contemporary digital revolution. If it is implemented correctly, we will reap its benefits for years to come without toil and hardship.

It is also necessary to enact a set of laws and regulations that govern the work of these tools and technologies. As is known, many of these tools work automatically. Once you enter a set of data, they give you ready-made outputs. The question that arises here is who is responsible for these results if they are incorrect or conflict with the highest interests of the work of those institutions or contradict their laws and regulations.

Here, another problem of using advanced artificial intelligence technologies appears, which falls under the title of "ethics of using artificial intelligence". If the door is left wide open without strict government control, we will move beyond personal blackmail to high-level government blackmail that may affect the security of the state. Therefore, it is incumbent upon systems to enact these laws and clarify the extent to which this technology can be used for fear of harming others. (**Erkkilä, 2023**).

### 3. A Case Study of Iraq and the Acceptance of AI Technologies by its Institutions

According to the Oxford Insights index, Iraq is considered one of the countries with medium capabilities in the field of artificial intelligence adoption. A quick look at the Iraq Country Index (33.40) and comparing it to the indicators of other countries within its geographic region in the Middle East or the Arab world, we conclude that there is a possibility of updating the infrastructure in all sectors, not just the education sector, which relies on AI programs and outputs.

Iraq's ranking of 133 out of 194 countries included in the aforementioned index means that we are ahead of about 61 countries with their own political and economic capabilities and scientific weight in the past and present. We cannot ignore the history of education in all those countries or belittle their rights. They all have a huge cultural and intellectual heritage.

Looking at the index from another perspective, and according to the data of each of the three government sectors adopted by the index, namely (government, technology, data, and infrastructure), the index points (33.40) were distributed as shown in the figure below:

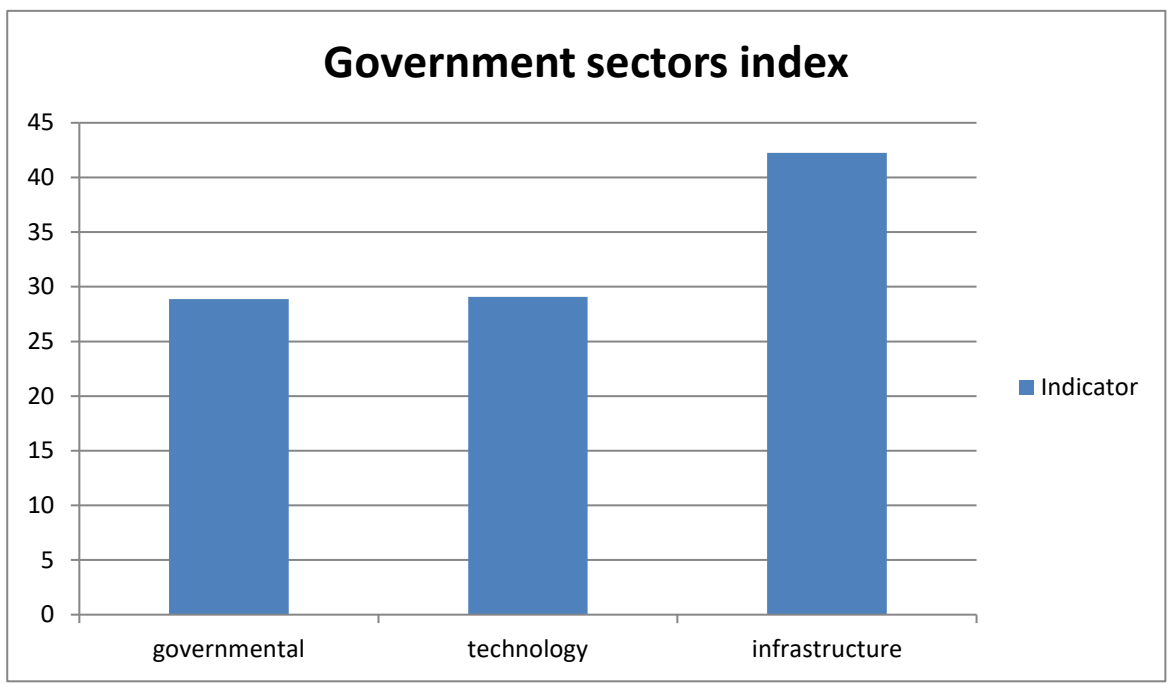

It is noticeable that the infrastructure and digital data liberation sector has surpassed the other two sectors (government sector and technology sector). This is very natural. Iraq needs more government capabilities and to intensify and continue efforts to develop the skills of workers in its government sectors at the level of decision-makers. I mean that the Iraqi government has not yet activated the digital government options except for very small percentages, and it is almost limited to complaints and presenting models of service transactions under the name of (e-citizen government). In the cases of submitting appointment and employment files, the government also activated the digital options. In some matters related to public services provided to the citizen, such as applying for a passport and other matters that fall under the category of public services, they were not at the required level.

In order for the Iraqi government to obtain high ranks in the scale of the global or regional index at least, it must activate all technical and technological options. It must update its digital systems, including communication networks, electronic sites, government communication accounts, and other security software. We must admit that the government digital transformation has become an important factor and a competitive element between the governments of the world, and these digital services have become the deciding factor for the fate of governments.

The recent period has witnessed a great interest in the digital revolution and artificial intelligence tools. The adoption of the vocabulary of this technology has become one of the most prominent features of modern learning. The relevant circles have been interested in updating and developing all sectors of this technology, which has been enhanced by the emergence of big data. This cognitive system has become appealing and attractive to all educational fields. (**Al-Attal, et al., 2021**). This technology has faced a number of problems and obstacles that have hindered the way of using its vocabulary.

One of the problems that hinder the integration of the smart digital system in Iraq is that internet services have not yet reached many remote areas of Iraq. Even the services provided in most urban areas of Iraq are not sufficient and are not at all effective. The networks are very weak. Yes, there is limited access to technology, and this is one of the most prominent challenges facing the Iraqi government. So how can it aspire to use artificial intelligence tools that require huge technical capabilities?

We will review what came in the United Nations Development Report for 2023 regarding the (digital landscape of Iraq), which relied on the analysis of (16) planning documents. The report shows that about 37% of the digital goals related to information and communication technology have been covered, with an assessment of 65 indicators out of 131 indicators. (**United Nations Development Program, 2023**). This is a weak percentage. The Iraqi government is trying to intensify its development efforts in line with the aspirations of international organizations, headed by the United Nations, which has presented, through its representatives in Iraq, a set of proposals to enhance digital integration, which is a fundamental part of the smart technology revolution and its tools.

Perhaps the most prominent of those recommendations presented in the aforementioned report are:

1- Long-term digital government strategy.
2- Digital government management and coordination

3. Systematic legal analysis

4. Consistency of public ICT funding

5. Digital data and interoperability

6. Cyber security

7. Access to ICT infrastructure

8. Identity and access to services

9. Digital skills

Each of these points needs a separate study, and at the same time, most of them are obstacles to the development of smart technical skills in Iraq that we are talking about.

If we focus on a small part of the components of digital government in Iraq, such as cybersecurity, for example, we will find that there are obstacles and difficulties that hinder the development of this aspect.

One of the obstacles is the lack of knowledge and understanding of the capabilities of artificial intelligence in the field of cybersecurity. There is also a lack of skills related to artificial intelligence, as well as the difficulty of providing a human infrastructure capable of responding to the skills of artificial intelligence tools in the field of cybersecurity. We should not forget the weak infrastructure in the institutions concerned with aspects of Iraqi digital security in terms of collecting, employing, analyzing, programming, and receiving its outputs.

Perhaps one of the most prominent points that everyone agrees on, and through which Iraq can gradually rise in the global and regional index of adoption of artificial intelligence technologies, is:

1. Enacting laws that govern the use of artificial intelligence tools and the handling of their outputs in all their forms and shapes.

2. Updating the infrastructure and human resources of employees and workers in all sectors of the state to keep pace with the artificial intelligence revolution. This can be done by involving young people in workshops, lectures, seminars, and conferences that focus on explaining and dealing with artificial intelligence. This is not difficult or impossible. Today, the world lives in a small digital village, and there are thousands of courses and seminars that are held virtually on the internet. It is also possible for all government sectors to contract with a group of specialists in this field to conduct these workshops for their employees on a regular basis.

3. The greatest burden lies on the educational and pedagogical ministries (the Ministry of Education and the Ministry of Higher Education), especially the latter. These ministries are responsible for developing training programs and schedules for their implementation by professional specialized professors. These courses should be disseminated to all ministries, especially those related to and interested in artificial intelligence, such as the Ministry of Interior, the Ministry of Defense, the Ministry of Science and Technology, the Ministry of Labor and Social Affairs, especially with regard to the provision of digital services to citizens, and the Ministry of Finance, which deals directly with banks, ATMs, and bank deposits.

4. Establishing the foundations for building advanced digital technical bases to ensure that inputs and knowledge materials are not lost, damaged, or hacked. This requires hiring professional and specialized people to build these bases and who are fully aware of the aspects of electronic hacking and to develop immediate and quick solutions to deal with such cases in a timely manner. It is also possible to rely on specialized technology companies in this field to manage the data entered into artificial intelligence software and applications used by the government.( Alalaq,2025)

5. Putting in place alternative and urgent plans in case any of these government websites that contain sensitive information are hacked, to ensure that this information is not leaked and that government or personal interests are not lost or blackmailed, and to ensure the preservation of privacy. Here, it is necessary to develop long-term plans, such as five-year plans that set visions and concepts for the near and distant future. It is known that any work without prior planning is not implemented in the correct practical way based on real data.

In the end, it can be said that artificial intelligence has become an integral part of our daily lives and has entered all sectors of life. It has become a reality that we must acknowledge, whether we like it or not. One of the most prominent manifestations of this reality is that everyone is now singing the praises of the benefits of artificial intelligence tools and is working day and night to update the infrastructure of these applications and bring them to the forefront of the world. The competition between research centers and specialists is fierce, which has produced a huge number of these programs and tools with a variety of tasks and functions. Some of them deal with security, others with the military, and others with finance, trade, education, guidance, services, and much more.

**Conclusion**:

In light of the increasing pace of work and the growing interest in the technologies of the contemporary digital revolution, one of the most prominent outputs and amazing results of which are artificial intelligence tools, widespread interests have emerged at the level of world countries from governments, educational and educational institutions, technology companies, and research centers. They are competing with each other to update the rules of these programs in a way that matches the labor market.

Recently, one of the most prominent and important indicators of interest in technology has emerged, monitoring the commitment of countries of the world to this technology and the extent to which it benefits from it in all aspects, namely the British (Oxford Insights) index. The index has included fixed criteria that it has followed for the past years to assess the readiness of the world's countries to adopt and deal with artificial intelligence tools and the extent to which they benefit from it in their government institutions. The index included three main sectors: (government sector, technology, data and infrastructure).( Alalaq,2024).

We obliged ourselves in this research to conduct an initial reading of the contents of the index while measuring the status of the Middle East and North Africa region, specifically the case of Iraq, and we came out with the result that the dominance of developed countries and Europe at the top of the pyramid of the index was and still is a natural result of the extent of those countries' commitment to fixed standards in their dealings with the developments of the digital revolution. This is due to the political power and the huge economic capabilities that these countries enjoy, which has a positive impact on the diversification of the knowledge chain of modern software programs.

In contrast to other countries, including the Middle East countries, we found that they occupy low ranks in the index series with some exceptions. We also identified the reasons for the hesitation of these countries to work and respond to artificial intelligence technology, the most prominent of which is the lack of financial resources allocated to update the infrastructure that deals with the digital revolution and its outputs. There was also a shortage of specialists and researchers in this new and emerging field. Therefore, if these countries want to catch up with the developed European countries, it is necessary to race against time, accelerate the pace, and put in place future plans in order to have a role in the field of digital technologies.